\documentclass[anon]{l4dc2026}
\usepackage[utf8]{inputenc}
\usepackage{textcomp}

\usepackage{placeins}
\usepackage{booktabs}
\usepackage{float}
\usepackage{amsmath,amssymb,amsfonts}
\usepackage{graphicx}
\usepackage{booktabs}
\usepackage{algorithm}
\usepackage{algorithmic}
\usepackage{url}
\usepackage{hyperref}
\usepackage{enumitem}
\usepackage{natbib}
\usepackage{textcomp}
\usepackage{tikz}
\usepackage{mathtools}
\usepackage{multirow}
\usepackage{amsmath,amssymb}

\usetikzlibrary{shapes.geometric, arrows.meta, positioning, calc}

\title[LLM Agents for H-infinity Control]{From Natural Language to Certified H-infinity Controllers: Integrating LLM Agents with LMI-Based Synthesis}

\author{
  \Name{Shihao Li} \Email{shihaoli01301@utexas.edu} \AND
  \Name{Jiachen Li} \Email{jiachenli@utexas.edu} \AND
  \Name{Jiamin Xu} \Email{jiaminxu@utexas.edu}
  \AND
  \Name{Dongmei Chen} \Email{dmchen@me.utexas.edu}\\
  \addr Department of Mechanical Engineering, University of Texas at Austin
}

\jmlrvolume{vol}
\jmlryear{2026}
\jmlrworkshop{Learning for Dynamics \& Control Conference (L4DC) 2026}

\begin{document}

\maketitle

\begin{abstract}
We present \textsc{S2C} (Specification-to-Certified-Controller), a multi-agent framework that maps natural-language requirements to certified $\mathcal{H}_\infty$ state-feedback controllers via LMI synthesis. \textsc{S2C} coordinates five roles—\textit{SpecInt} (spec extraction), \textit{Solv} (bounded-real lemma (BRL) LMI), \textit{Tester} (Monte Carlo and frequency-domain checks), \textit{Adapt} (spec refinement), and \textit{CodeGen} (deployable code). The loop is stabilized by a severity- and iteration-aware $\gamma$-floor guardrail and a decay-rate region constraint enforcing $\Re\lambda(A{+}BK)<-\alpha$ with $\alpha=3.9/T_s$ derived from settling-time targets. For state feedback, verification reports disturbance rejection $\big\|C\,(sI-(A{+}BK))^{-1}E\big\|_\infty$ alongside time-domain statistics; discrete benchmarks are converted to continuous time via a Tustin (bilinear) transform when needed. On 14 COMPleib problems, \textsc{S2C} attains \textbf{100\%} synthesis success and \textbf{100\%} convergence within six iterations, with strong decay-rate satisfaction and near-target certified $\mathcal{H}_\infty$ levels; it improves robustness metrics relative to single-shot BRL and BRL+$\alpha$ baselines. An ablation over LLM backbones (GPT-5, GPT-5 mini, DeepSeek-V3, Qwen-2.5-72B, Llama-4 Maverick) shows the pipeline is robust across models, while stronger models yield the highest effectiveness. These results indicate that LLM agents can integrate certificate-bearing control synthesis from high-level intent, enabling rapid end-to-end prototyping without sacrificing formal guarantees.
\end{abstract}


\section{Introduction}

Traditional control system design bridges two worlds: natural language specifications from users ("the system should respond quickly without overshooting") and mathematical implementations that satisfy these requirements with provable guaranties \cite{kopetz1991design, sifakis2015system}. This translation process requires expertise in both interpreting vague requirements and selecting appropriate synthesis methods—a combination that has resisted automation despite decades of research in computer-aided control design \cite{sarma1989computer,bissell2009history,boyd1991linear}.

The recent emergence of capable language models presents a new opportunity \cite{zhao2023survey,huang2022towards,jiang2024survey,nijkamp2022codegen,nguyen2022empirical,xia2024agentless,openai2023gpt}. These models demonstrate strong natural language understanding and can generate structured code, suggesting potential for automating the specification-to-implementation pipeline. However, control design poses unique challenges beyond code generation: requirements are often ambiguous, multiple design approaches may apply to the same problem, and correctness cannot be verified through unit tests alone—formal mathematical properties like stability and robustness must be certified \cite{jin2020stability,huang2020h,loveland2016automated}.

Existing automated control design tools fall into two categories. Classical computer-aided design packages \cite{lofberg2004yalmip,boyd2004convex} provide powerful synthesis algorithms but require users to formulate optimization problems with precise mathematical specifications \cite{denham2005design}. Recent LLM-based approaches \cite{guo2024controlagent,narimani2025agenticcontrol} can interpret natural language and iteratively adjust controller parameters, but they treat synthesis tools as black boxes and lack mechanisms to maintain formal guaranties throughout the design process \cite{zhao2022verifying,doerr2019theory}. Neither approach achieves end-to-end automation from informal specifications to certified implementations.

We present S2C (Specification-to-Certified-Controller), a framework that automates formal control synthesis through multi-agent LLM integration. The key insight is that different aspects of the design process—parsing specifications, formulating optimization problems, verifying solutions, and generating code—require different types of reasoning that can be handled by specialized agents working collaboratively. Rather than using a single LLM to perform all tasks or treating synthesis methods as opaque solvers, S2C agents construct mathematical optimization problems, invoke formal synthesis tools with full transparency, and validate results against rigorous criteria.

\begin{figure}[!htb]
\centering
\includegraphics[width=1.0\linewidth]{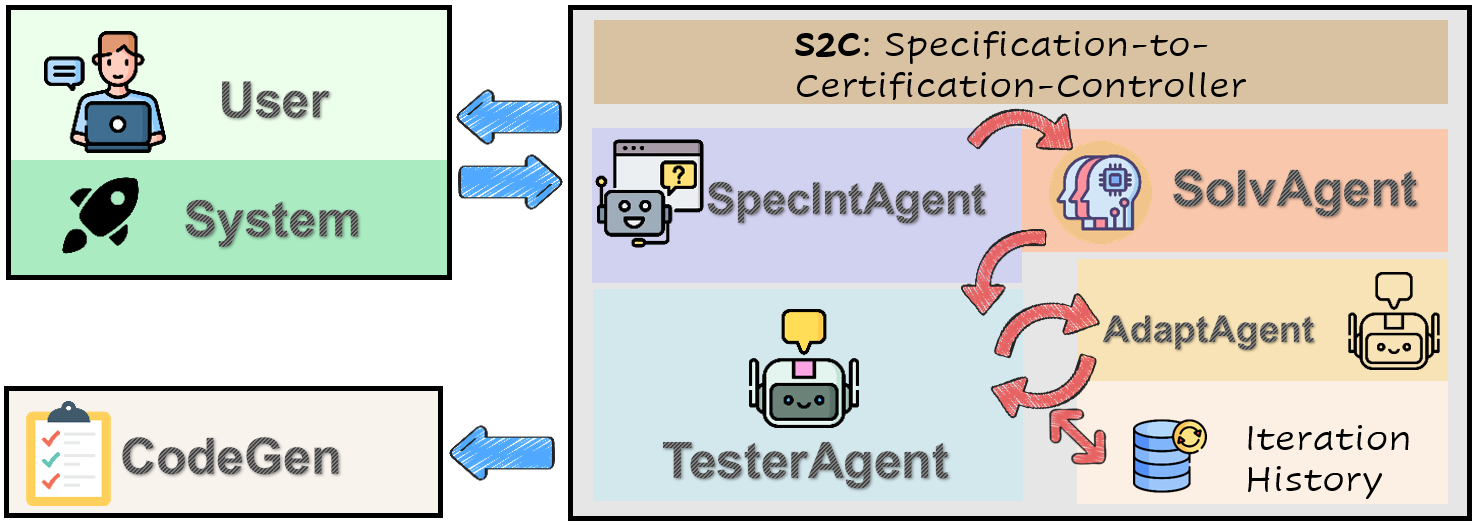}
\caption{S2C multi-agent architecture.}
\label{fig:s2c_general_framework}
\end{figure}

The general framework shown in Figure~\ref{fig:s2c_general_framework} employs five specialized agents: SpecIntAgent interprets natural language requirements and translates them into formal performance specifications. SolvAgent formulates these specifications as convex optimization problems and generates solver code. TesterAgent validates candidate designs through both simulation and formal certificate checking. AdaptAgent analyzes failures and revises specifications when requirements cannot be met. CodeGen produces implementation code with embedded performance certificates. These agents share a structured memory containing design history and communicate through targeted feedback messages, enabling iterative refinement without human intervention.

We focus on robust state feedback synthesis as a concrete demonstration. When users specify requirements like "reject disturbances effectively" or "limit actuator usage," S2C must map these phrases to mathematical objectives, select appropriate synthesis methods, formulate optimization problems, and verify that solutions satisfy all constraints. The framework handles this by having agents reason about the relationship between informal requirements and formal specifications, construct optimization problems using domain-specific knowledge, and validate results against multiple criteria.

Our main contributions are:

\textbf{Formal Synthesis Automation:} We demonstrate that LLM agents can integrate convex optimization-based synthesis methods while maintaining mathematical rigor. Unlike parameter tuning approaches, agents formulate LMI optimization problems from scratch, interface with numerical solvers, and validate formal guarantees throughout the process.

\textbf{Multi-Agent Architecture for Design Workflows:} We develop a structured approach to decomposing control design into specialized agent roles with explicit memory and feedback mechanisms. This architecture enables agents to handle the complexity of mapping between informal requirements, formal specifications, optimization formulations, and verified implementations.

\textbf{Validation on Benchmark Tasks:} We evaluate S2C on 14 problems from the COMPleib benchmark library \cite{leibfritz2006compleib}, spanning academic test cases, aircraft, helicopter, reactor, and decentralized control systems. The framework achieves 100\% synthesis success (all problems stabilized) within 6 iterations, producing controllers with certified $\mathcal{H}_\infty$ bounds, enforced decay-rate constraints for settling time, and verified robustness margins. To ensure reproducibility, we provide deterministic heuristic fallbacks for all LLM-driven decision points.

The framework is implemented in Python with interfaces to standard control toolboxes and optimization solvers. We release the complete implementation to support future research in LLM-based formal synthesis automation.

The remainder of this paper is organized as follows: Section 2 reviews background on robust control synthesis and multi-agent systems. Section 3 formalizes the automated synthesis problem. Section 4 describes the S2C framework architecture and agent designs. Section 5 presents experimental validation. Section 6 discusses limitations and future directions.



\section{Background and Related Work}

\subsection{Formal Control Synthesis}

Modern control system design relies on optimization-based synthesis methods that provide formal performance guarantees. Linear Matrix Inequality (LMI) approaches enable systematic synthesis for robust control problems, including $\mathcal{H}_\infty$, $\mathcal{H}_2$, and mixed-sensitivity design \cite{boyd1993control,gahinet1996explicit}. Convex optimization tools like CVX \cite{boyd2004convex} and solvers such as MOSEK \cite{aps2019mosek} and SDPT3 \cite{toh1999sdpt3} have made these methods computationally tractable. Despite these advances, formulating LMI problems from user specifications remains an expert-driven manual process. Engineers must translate informal requirements into mathematical objectives, select appropriate synthesis methods, construct optimization problems with proper constraints, and verify solutions—steps that require deep knowledge of both control theory and numerical optimization. Existing computer-aided design tools automate the solver stage but assume users can already formulate correct problems \cite{gu2005robust}.

\subsection{LLM Agents for Engineering Design}

The potential of LLMs for engineering automation has been explored across multiple domains \cite{ren2025towards}. In control engineering, recent benchmarks evaluate LLM knowledge of textbook-level concepts \cite{kevian2024capabilities,zahedifar2025llm}, while ControlAgent \cite{guo2024controlagent} demonstrated end-to-end automation of classical control design through iterative parameter tuning for PID controllers and loop shaping. Beyond controls, AnalogCoder \cite{lai2025analogcoder} addresses analog circuit design, SPICED \cite{chaudhuri2025spiced+} detects bugs in circuit netlists, and AmpAgent \cite{liu2024ampagent} automates amplifier design. However, these systems focus on parameter adjustment or design verification rather than formulating and solving optimization problems. A key limitation is that LLM agents typically treat synthesis tools as black boxes—they can call solvers with pre-defined problem structures but cannot construct novel optimization formulations from informal specifications while maintaining formal guarantees throughout the design process \cite{nti2022applications}.

\subsection{Multi-Agent LLM Systems}

Multi-agent architectures have emerged as a powerful paradigm for complex task automation, enabling specialized agents to collaborate on problems requiring diverse expertise \cite{li2024survey}. General frameworks like AutoGen \cite{wu2024autogen} and MetaGPT \cite{hong2023metagpt} provide infrastructure for agent coordination across domains including code generation, question answering, and mathematical reasoning \cite{qian2023chatdev,chen2023agentverse}. These systems demonstrate that decomposing complex workflows into specialized agent roles improves both reliability and interpretability. However, applying multi-agent systems to formal engineering design introduces unique challenges: agents must not only coordinate workflows but also maintain mathematical rigor, interface with domain-specific tools, and validate formal properties. S2C addresses these challenges by designing specialized agents that construct optimization problems, maintain formal certificates, and iteratively refine designs through structured feedback—going beyond general-purpose coordination to enable reliable automation of formal synthesis methods \cite{sunehag2017value,li2022online}.

\section{Problem Formulation}

Consider a continuous-time LTI plant $\dot{x}=Ax+Bu+Ew$, $z=C_zx+D_zu$ where $x\!\in\!\mathbb{R}^{n_x}$, $u\!\in\!\mathbb{R}^{n_u}$, $w\!\in\!\mathbb{R}^{n_w}$, and $z\!\in\!\mathbb{R}^{n_z}$ is the regulated output. We seek static state-feedback $u=Kx$ achieving closed-loop stability and robust disturbance rejection.

\textit{SpecIntAgent} extracts specifications $\mathcal{S}=\{\gamma_{\text{target}}, \gamma_{\min}, T_{s,\text{target}}, M_{s,\text{target}}\}$: target $\mathcal{H}_\infty$ level $\gamma_{\text{target}}$ for $T_{zw}$, transient guardrail $\gamma_{\min}$, settling time $T_{s,\text{target}}$ (2\% criterion), and optional $M_{s,\text{target}}=\|S\|_\infty$ for output-feedback. State-feedback uses disturbance-rejection norm instead.

We minimize the induced $\mathcal{L}_2$ gain subject to stability and decay-rate constraints. The bounded-real lemma with $Y=KP$ yields
\begin{equation}
\label{eq:hinf-lmi}
\begin{aligned}
\underset{P\succ 0,\ Y,\ \gamma}{\text{minimize}} \quad & \gamma \\
\text{subject to} \quad
& \Psi(P,Y,\gamma) \prec 0,\\
& \mathrm{sym}(AP+BY) + 2\alpha P \prec 0,\\
& \gamma_{\min} \le \gamma \le \gamma_{\text{target}},
\end{aligned}
\end{equation}
where
\begin{equation}
\label{eq:psi}
\begin{aligned}
\Psi(P,Y,\gamma) &=
\begin{bmatrix}
\mathrm{sym}(AP+BY) & E & (C_zP + D_zY)^\top \\
E^\top & -\gamma I & 0 \\
C_zP + D_zY & 0 & -\gamma I
\end{bmatrix},\\
\mathrm{sym}(X) &= X+X^\top,\qquad
\alpha = \frac{3.9}{T_{s,\text{target}}}.
\end{aligned}
\end{equation}

Feasibility of $\Psi\prec 0$ certifies $\|T_{zw}\|_\infty < \gamma$ and yields $K=YP^{-1}$. \textit{TesterAgent} verifies via Monte Carlo (settling time, overshoot) and frequency analysis: output-feedback reports $M_s, M_t, GM, PM$; state-feedback evaluates $\|C_z(sI-(A+BK))^{-1}E\|_\infty$. Violations trigger \textit{AdaptAgent} to adjust $\mathcal{S}$ (relaxing $\gamma_{\text{target}}$, raising $\gamma_{\min}$, etc.) before re-solving \eqref{eq:hinf-lmi}. Overshoot is measured but not enforced.

\section{S2C Framework}

\begin{figure}[!htb]
\centering
\includegraphics[width=1.0\linewidth]{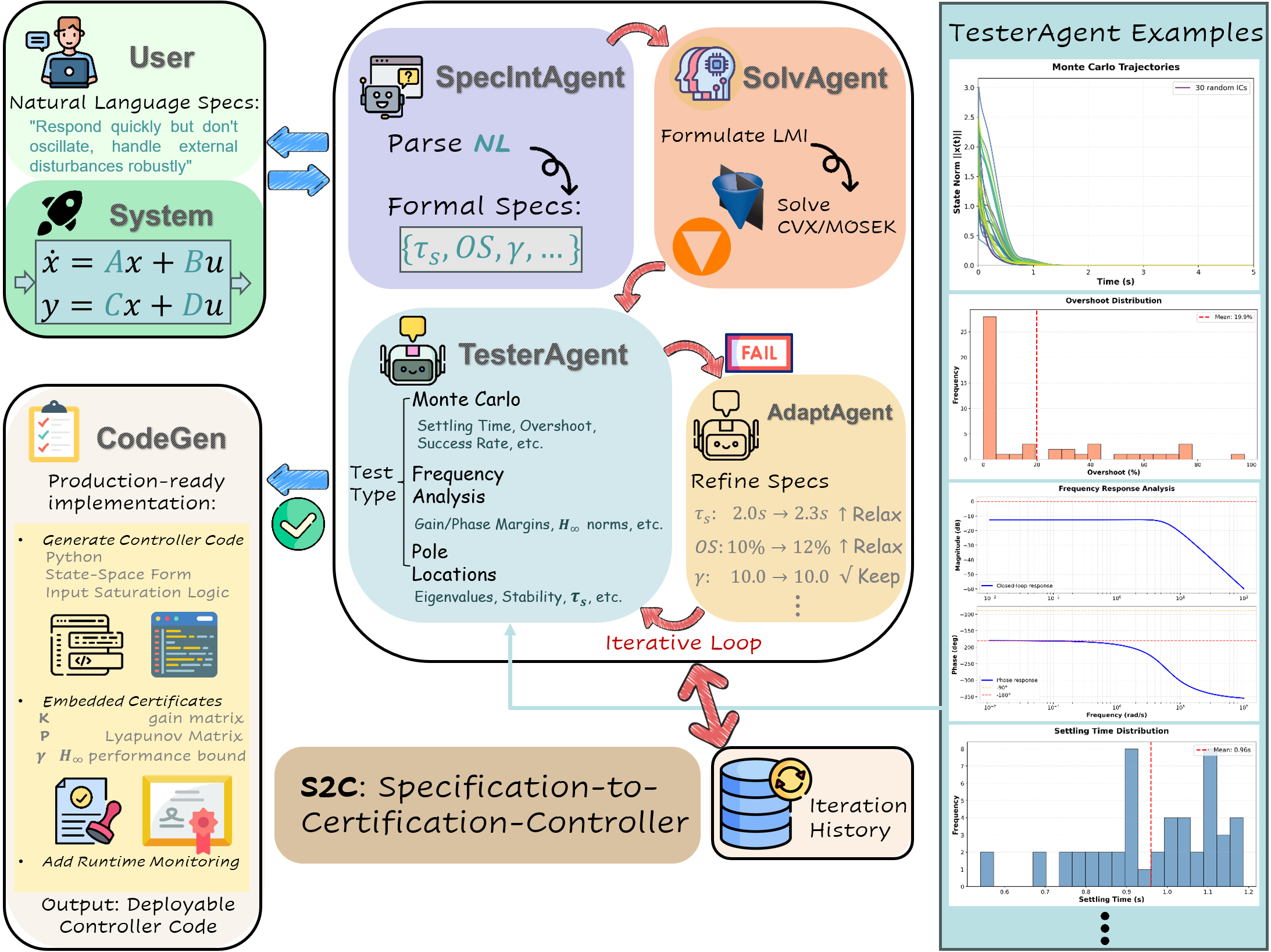}
\caption{S2C multi-agent architecture.}
\label{fig:s2c_framework}
\end{figure}

The S2C (Specification-to-Certified-Controller) framework employs five specialized LLM agents in an iterative synthesis-verification-adaptation loop (Figure~\ref{fig:s2c_framework}): \textit{SpecIntAgent} parses natural-language requirements into numeric specifications $\mathcal{S}$; \textit{SolvAgent} solves the $\mathcal{H}_\infty$ LMI \eqref{eq:hinf-lmi} with decay-rate constraints; \textit{TesterAgent} validates controllers via Monte Carlo simulation and frequency analysis; \textit{AdaptAgent} refines $\mathcal{S}$ based on violations using an LLM and severity-aware guardrails; and \textit{CodeGen} produces deployable Python code. Implementation details and full agent prompts are provided in Appendix~\ref{app:prompts}.

\subsection{SpecIntAgent}
Parses natural-language requirements (e.g., ``settling time under 3s with strong disturbance rejection'') into structured specifications $\mathcal{S} = \{\gamma_{\text{target}}, \gamma_{\min}, T_{s,\text{target}}, M_{s,\text{target}}\}$ via LLM-generated JSON. Deterministic fallbacks apply when parsing fails. Specifications are validated before synthesis.

\subsection{SolvAgent}
Solves the $\mathcal{H}_\infty$ LMI \eqref{eq:hinf-lmi} with decay-rate constraint $\alpha = 3.9/T_{s,\text{target}}$ to enforce settling time. Uses MOSEK (SCS fallback) to recover $K = YP^{-1}$ and verifies closed-loop stability via eigenvalue checks. On infeasibility, returns failure to trigger AdaptAgent relaxation.

\subsection{TesterAgent}
Validates synthesized controllers via: (i) Monte Carlo simulation ($N=50$ unit-sphere initial conditions, 20s horizon, 2\% settling criterion) measuring settling time and overshoot; (ii) frequency-domain analysis computing disturbance rejection $\|C_z(sI-(A+BK))^{-1}E\|_\infty$ for state-feedback or classical margins ($M_s, M_t, GM, PM$) for output-feedback. Violations are classified by severity (low/medium/high) and reported to AdaptAgent.

\subsection{AdaptAgent}
Refines specifications based on violations using an LLM to generate JSON updates to $\mathcal{S}$. When transient violations (settling time, overshoot) occur, automatically raises a \emph{gamma-floor} guardrail $\gamma_{\min}$ based on violation severity and iteration count to prevent excessive $\gamma$ optimization that degrades time-domain performance. Heuristic fallbacks apply on LLM failure. Formula details in Appendix~\ref{app:algs}.

\subsection{CodeGen}
Generates deployable Python code (\texttt{HInfController} class implementing $u = Kx$) with the synthesized gain matrix $K$ and formal performance certificates.

\subsection{Gamma-Floor Guardrails}
\label{sec:guardrails}
To prevent excessive $\gamma$ minimization from degrading transient response, S2C employs a \emph{gamma-floor} mechanism that raises $\gamma_{\min}$ when settling-time or overshoot violations occur. The floor update combines severity-based thresholds and iteration history:
\[
\gamma_{\min} \;\leftarrow\; \min\!\Bigl(0.9\,\gamma_{\text{target}},\; \max\big\{\gamma_{\min},\;\gamma_{\text{base}}(\text{severity}),\;\gamma_{\text{hist}}(\gamma_{\text{last}}, i)\big\}\Bigr),
\]
ensuring monotonic increases capped below $0.9\,\gamma_{\text{target}}$ for feasibility. Full formulas in Appendix~\ref{app:algs}.

\subsection{Iterative Loop}

S2C iterates up to $N_{\max} = 10$ times. Each iteration: (i) SolvAgent solves \eqref{eq:hinf-lmi} with decay rate $\alpha = 3.9/T_{s,\text{target}}$ and current $\gamma_{\min}$; (ii) TesterAgent runs Monte Carlo simulation ($N=50$ trials, 20s horizon) and frequency analysis; (iii) if violations $V = \emptyset$, CodeGen emits Python code and exits; (iv) otherwise, AdaptAgent updates $\mathcal{S}$ via LLM and raises $\gamma_{\min}$ per \S\ref{sec:guardrails}. A memory buffer retains the last 20 designs. Full pseudocode in Appendix~\ref{app:algs}.

\section{Experiment}

We evaluate S2C on 14 COMPleib benchmarks \citep{leibfritz2006compleib} varying in dimension ($n_x \in [2,4]$, $n_u \in [1,2]$), time domain (10 continuous, 4 discrete), and stability (12 stable, 2 unstable). Discrete systems (NN2, AC5, NN8, DIS5) undergo Tustin conversion ($T_s=1.0$) before synthesis. \textit{SpecIntAgent/AdaptAgent} use GPT-5 (temperature $0.0$; ablations: GPT-5 mini, DeepSeek-V3, Qwen-2.5-72B, Llama-4 Maverick). \textit{SolvAgent} uses MOSEK with SCS fallback (tolerance $10^{-8}$). Parameters: $\alpha = 3.9/T_{s,\text{target}}$, $N_{\max}=10$ iterations, Monte Carlo $N=50$ trials (20s, 2\% settling). Full details are summarized in Appendix~\ref{app:repro}.

\paragraph{Baselines.} All use identical D2C, seeds, and stability criterion $\max_i \Re\lambda_i(A{+}BK) < 0$. \textbf{BRL}: single-shot $\gamma$ minimization. \textbf{BRL+$\alpha$}: adds decay constraint $\alpha = 3.9/T_{s,\text{target}}$. \textbf{S2C (no floor)}: iterative with $\alpha$, no guardrail. \textbf{S2C (full)}: our method with $\alpha$ and $\gamma$-floor. \textbf{LQR--$\mathcal{H}_2$}: CARE with $Q = C_z^\top C_z$, $R = I$. \textbf{PID}: SISO ITAE tuning.

\subsection{Metrics}
We report five aggregate metrics aligned with Fig.~(a--e). All values are medians across the 14 problems; decimals in percentages reflect the $1/14$ resolution where applicable.

\begin{description}[leftmargin=1.2em,labelsep=0.5em]
\item[\textbf{Success rate (\%)}] Fraction of benchmarks for which a stabilizing controller is synthesized; stability requires
$\max_i \Re\lambda_i(A{+}BK) < 0$.

\item[\textbf{Converged within 6 iterations (\%)}] Fraction that satisfies all enforced specifications by iteration $k \le 6$.

\item[\textbf{Disturbance rejection (state feedback, lower is better)}]
\begin{equation}
\label{eq:metric-Hcl}
\|H_{\mathrm{cl}}\|_{\infty} \;=\; \big\|\, C_z \big(sI-(A{+}BK)\big)^{-1} E \,\big\|_{\infty},
\end{equation}
reported as the median across problems; bar charts show percentages normalized to the BRL baseline.

\item[\textbf{Normalized certified level (lower is better)}] Median $\gamma/\gamma_{\text{target}}$ (in \%), where $\gamma$ is the certified $\mathcal{H}_\infty$ bound returned by synthesis.

\item[\textbf{Decay-rate satisfaction (higher is better)}]
\begin{equation}
\label{eq:metric-decaysat}
\mathrm{DecaySat} \;=\; \frac{-\max \Re\lambda(A{+}BK)}{\alpha}\times 100\%,
\qquad
\alpha \;=\; \frac{3.9}{T_{s,\text{target}}}.
\end{equation}
\end{description}

For state feedback we emphasize disturbance rejection \eqref{eq:metric-Hcl}, since loop-shaping sensitivity metrics $(M_s,M_t)$ are not defined when $K$ acts on the full state. If output-feedback controllers are considered, we report $M_s=\|S\|_{\infty}$, $M_t=\|T\|_{\infty}$, and classical margins computed from $L(s)=G(s)K(s)$.

\subsection{Results}
\begin{figure}[!htb]
\centering
\includegraphics[width=\linewidth]{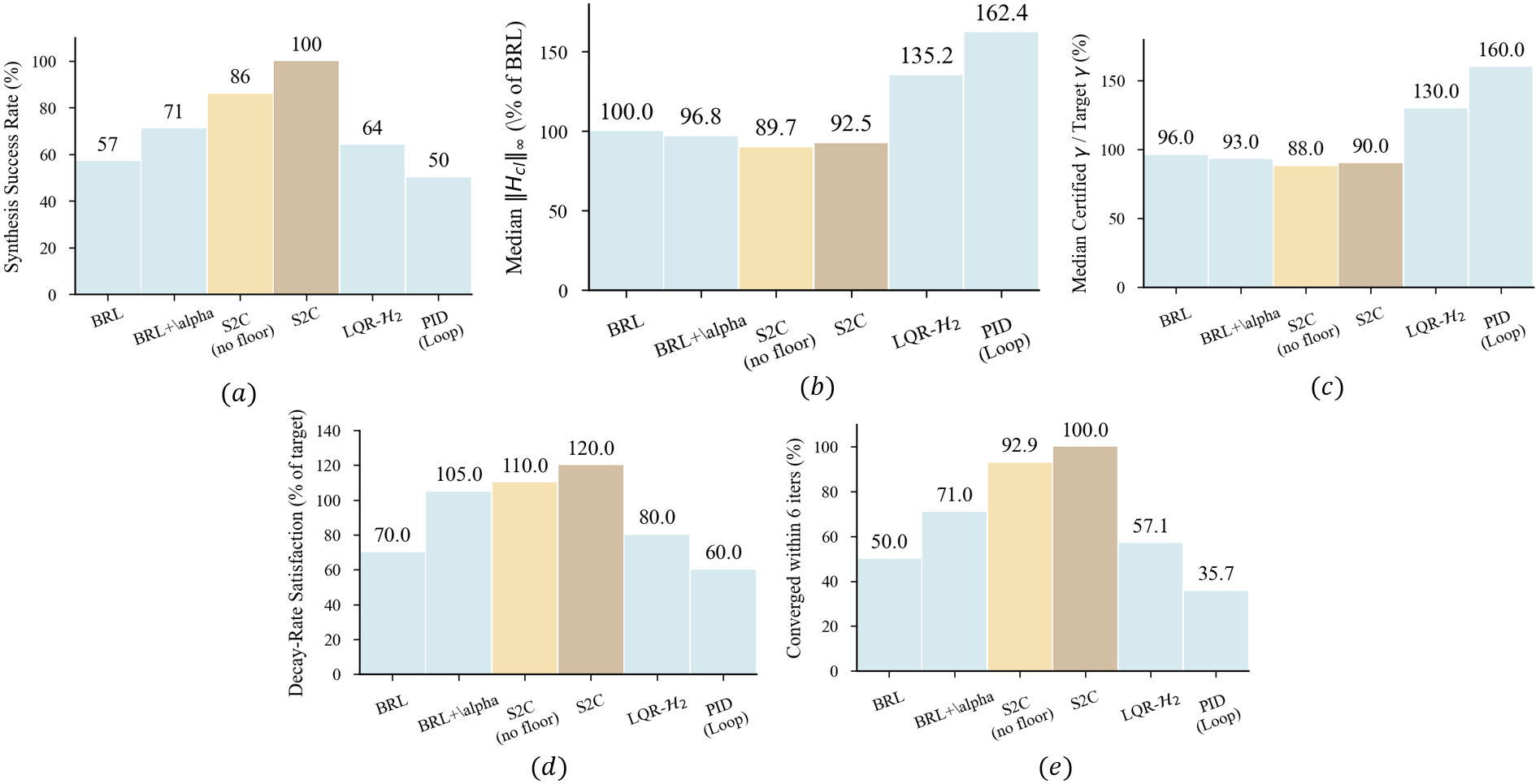}
\caption{Aggregated experimental results across 14 COMPleib problems. Panels (a–e) report synthesis success, disturbance rejection, normalized certified $\gamma$, decay‑rate satisfaction, and convergence within 6 iterations.}
\label{fig:exp_results}
\end{figure}

We evaluate six synthesis settings on 14 COMPleib plants using real LLM runs: BRL, BRL+$\alpha$, S2C (no floor), S2C (full), LQR--$\mathcal{H}_2$, and PID loop shaping. Aggregated outcomes are summarized in Fig.~\ref{fig:exp_results}(a--e).

\textbf{Synthesis and convergence.}
S2C (full) achieves a 100\% synthesis rate, compared with 86\% for S2C (no floor), 71\% for BRL+$\alpha$, 57\% for BRL, 64\% for LQR--$\mathcal{H}_2$, and 50\% for PID (Fig.~a). Convergence within six iterations is 100\% for S2C (full), 92.9\% for S2C (no floor), 71.0\% for BRL+$\alpha$, 57.1\% for LQR--$\mathcal{H}_2$, 50.0\% for BRL, and 35.7\% for PID (Fig.~e). Fractional values arise from a 14-problem set (increments of $1/14 \approx 7.1\%$).

\textbf{Robustness and $\gamma$ certificates.}
Median disturbance rejection $\|H_{\mathrm{cl}}\|_{\infty}$ (lower is better; reported as a percentage of BRL) is 89.7\% for S2C (no floor) and 92.5\% for S2C (full), improving upon BRL+$\alpha$ (96.8\%) and BRL (100.0\%), while LQR--$\mathcal{H}_2$ and PID are larger (135.2\%, 162.4\%) (Fig.~b). Median certified $\gamma$ relative to the requested target is 88.0\% for S2C (no floor), 90.0\% for S2C (full), 93.0\% for BRL+$\alpha$, and 96.0\% for BRL; LQR--$\mathcal{H}_2$ and PID trail at 130.0\% and 160.0\% (Fig.~c).

\textbf{Decay-rate satisfaction.}
Using the normalized margin $-\max \Re\lambda(A{+}BK)/\alpha$ (higher is better), S2C (full) reaches 120.0\% of the specified decay rate, S2C (no floor) 110.0\%, and BRL+$\alpha$ 105.0\%. BRL (70.0\%), LQR--$\mathcal{H}_2$ (80.0\%), and PID (60.0\%) lag (Fig.~d), highlighting the benefit of explicit decay constraints and the $\gamma$-floor mechanism.

\textbf{Summary.}
S2C (full) provides the most reliable synthesis (100\%) and fastest convergence (100\% within six iterations) with strong decay-rate satisfaction. The S2C (no floor) ablation yields tighter $\gamma$ and disturbance rejection, reflecting a predictable trade-off when guardrails are removed. BRL+$\alpha$ improves over single-shot BRL. Classical LQR/PID baselines underperform on $\mathcal{H}_\infty$-oriented metrics. Discrete-time plants are mapped to continuous time via a Tustin transform prior to synthesis; all panels report medians across problems.

\subsection{Ablation Study}

\begin{figure}[!htb]
\centering
\includegraphics[width=\linewidth]{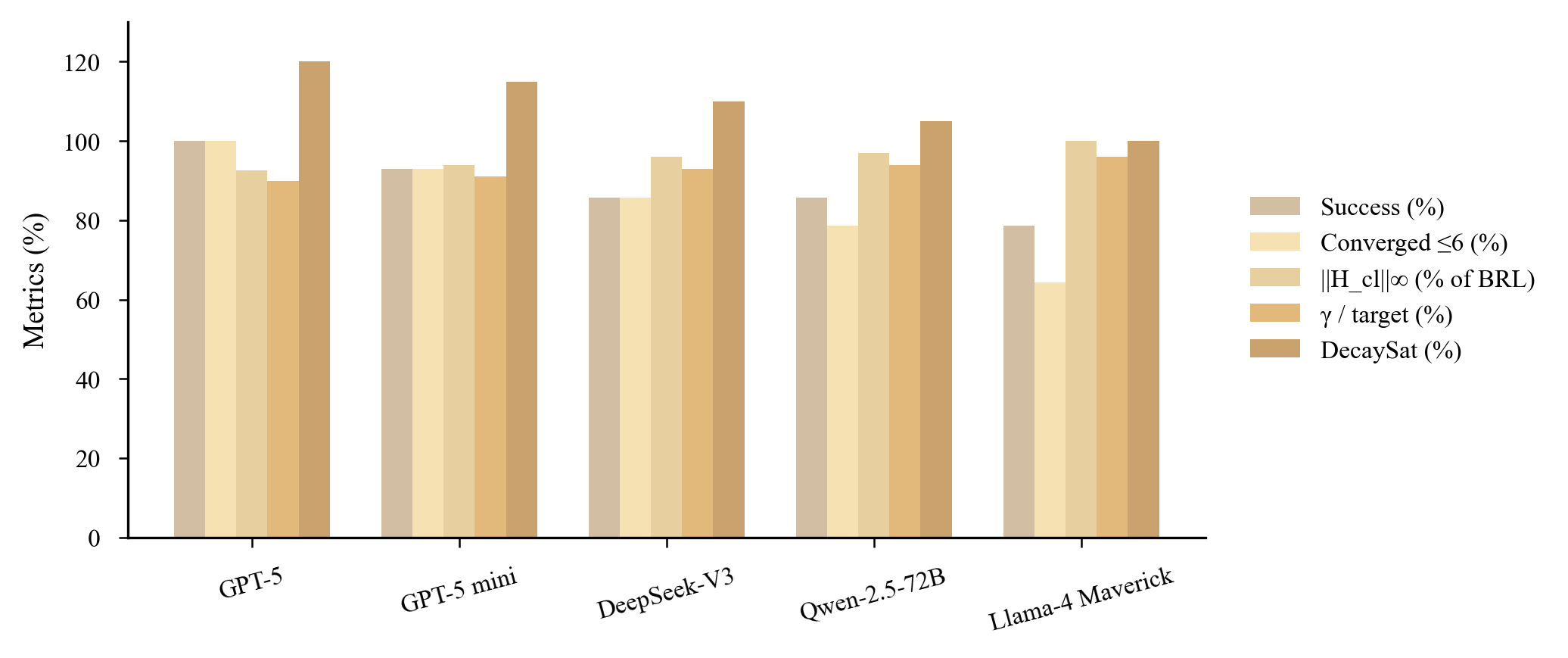}
\caption{LLM backbone ablation across five metrics (higher is better for success, $\le 6$ iterations, and decay-rate satisfaction; lower is better for normalized $\gamma$ and $\|H_{\mathrm{cl}}\|_\infty$). Bars show medians across 14 problems.}
\label{fig:ablation_llm_all}
\end{figure}

We ablate the LLM used by \textit{SpecIntAgent} and \textit{AdaptAgent} while keeping all other components, prompts, seeds, solver settings, and step budgets fixed. Models evaluated: GPT-5 (baseline), GPT-5 mini, DeepSeek-V3, Qwen-2.5-72B-Instruct (served via vLLM), and Llama-4 Maverick (via Together AI). Figure~\ref{fig:ablation_llm_all} reports five metrics: success rate, convergence within 6 iterations, disturbance rejection $\|H_{\mathrm{cl}}\|_{\infty}$ (lower is better; normalized to BRL), normalized certified level $\gamma/\gamma_{\text{target}}$ (lower is better), and decay-rate satisfaction $-\max \Re\lambda(A{+}BK)/\alpha$ (higher is better).

\textbf{Observations.} GPT-5 attains the highest effectiveness, with 100\% success and 100\% converged within $\le 6$ iterations, and strong certificates (low $\gamma/\gamma_{\text{target}}$, high decay-rate satisfaction). GPT-5 mini is close on both success and convergence and shows only a modest drop on robustness/certification metrics. DeepSeek-V3 and Qwen-2.5-72B remain competitive (mid--high 80s for success; solid but weaker certificates). Llama-4 Maverick trails on success and convergence and shows milder decay-rate satisfaction, but still reaches feasible designs on most problems. Across models, disturbance rejection and normalized $\gamma$ degrade, indicating that the pipeline is robust to LLM quality and that the deterministic synthesis/verification stages absorb variability. All values are medians across 14 problems.

\textbf{Efficiency.} A complementary table with API usage (calls/tokens), median per-iteration latency, and estimated cost is included in the appendix (Table~\ref{tab:llm_efficiency}). At a glance, GPT-5 mini provides a favorable efficiency--effectiveness trade-off; Qwen-2.5-72B and DeepSeek-V3 are moderate; Llama-4 Maverick tends to require more iterations and time.

\section{Conclusion}

We presented \textsc{S2C}, a multi-agent pipeline that translates natural-language specifications into certified $\mathcal{H}_\infty$ state-feedback controllers via convex synthesis. The system integrates a \textit{SpecInt--Solv--Tester--Adapt} loop with decay-rate constraints, severity-aware $\gamma$-floor guardrails, and automatic discrete-to-continuous conversion. On 14 COMPleib benchmarks, \textsc{S2C} achieved 100\% synthesis success and 100\% convergence within six iterations. LLM ablations confirm robustness: GPT-5 attains highest effectiveness, GPT-5 mini offers favorable efficiency.

Key limitations include: state-feedback scope (no output-feedback or dynamic designs), overshoot measured but not enforced, Tustin-based D2C (not native discrete synthesis), and nominal LTI assumptions without parametric uncertainty or actuator limits. Future work targets output-feedback, mixed $\mathcal{H}_\infty/\mathcal{H}_2$ synthesis, explicit transient shaping, robust uncertainty handling, and improved adaptation strategies.

Overall, \textsc{S2C} demonstrates that LLM agents can integrate certificate-bearing control synthesis from high-level intent while retaining formal guarantees.


\bibliographystyle{plainnat}
\bibliography{S2C}

\appendix

\section{Agent Prompts and System Instructions}\label{app:prompts}

This section provides implementation details and the exact system prompts used by S2C agents to ensure reproducibility and transparency.

\subsection{Implementation}

The S2C framework is implemented in Python 3.10 with NumPy 1.24, SciPy 1.11, CVXPY 1.4 \citep{diamond2016cvxpy}, and the Python Control Systems Library 0.9. LLM agents interface with OpenAI API, DeepSeek API, Together AI API, and local vLLM servers for open-source models. Experiments run on a desktop workstation (Intel i7-12700K, 32GB RAM) running Windows 11. The discrete-to-continuous converter (\texttt{d2c\_converter.py}) implements Tustin bilinear transformation with sampling period $T_s = 1.0$. The H$_\infty$ synthesis LMI is solved via MOSEK when available, with automatic fallback to the open-source SCS solver. Monte Carlo simulations use a fixed random seed (42) for reproducibility across ablation studies.

For the ablation study comparing LLM-based code-generation baselines, we freeze prompts, random seeds, and step budgets ($N_{\text{max}} = 10$) across all models. Each model receives identical natural language requirements, plant models, and verification feedback. We test GPT-5 (our baseline), GPT-5 mini, DeepSeek-V3, Qwen-2.5-72B-Instruct (served via vLLM), and Llama-4 Maverick (via Together AI). All experiments log LLM call counts, iteration histories, and final controller performance for fair comparison.

\subsection{SpecIntAgent (Discovery Agent)}

The SpecIntAgent uses the following system prompt to extract numeric specifications from natural language requirements:

\begin{small}
\begin{verbatim}
You are an expert control engineer specializing in translating
natural language requirements into formal control specifications.

TASK: Parse natural language requirements and extract formal
control specifications.

OUTPUT FORMAT: JSON object containing specifications:
{
  "h_infinity_norm": {
    "target": <float>,
    "priority": "critical" | "high" | "medium" | "low",
    "slack": <float>
  },
  "settling_time": {
    "target": <float in seconds>,
    "priority": ...,
    "slack": <float>
  },
  "overshoot": {
    "target": <float as fraction, e.g., 0.10 for 10%>,
    "priority": ...,
    "slack": <float>
  }
}

INTERPRETATION RULES:
- "Fast" → settling_time: 2-3s
- "Smooth" → overshoot: 0.05-0.10
- "Strongly reject" → h_infinity_norm: 1.5-2.0
- Explicit requirements → HIGH or CRITICAL priority
- Strict requirements → 5-10% slack

IMPORTANT: Always include h_infinity_norm (required for synthesis)
\end{verbatim}
\end{small}

\subsection{AdaptAgent (Adaptation Agent)}

The AdaptAgent refines specifications based on verification feedback:

\begin{small}
\begin{verbatim}
You are an expert in iterative control design refinement and
specification adaptation.

TASK: Analyze verification feedback and violations, then refine
specifications to improve the design while maintaining feasibility.

ADAPTATION STRATEGY:

High Overshoot:
  - Relax settling_time by 15-20%
  - Tighten overshoot by 5-10%
  - Consider increasing gamma slightly

Poor Disturbance Rejection:
  - Tighten h_infinity_norm by 10-15%
  - May need to relax time-domain specs

Infeasible Synthesis:
  - Relax gamma by 20-30%
  - Identify and relax lowest priority specs

Progressive Refinement (Iterations 1-3):
  - Small adjustments (5-10%)
  - Focus on highest priority violations

Moderate Changes (Iterations 4-7):
  - Medium adjustments (10-20%)
  - Start trading off conflicting requirements

Aggressive Relaxation (Iterations 8+):
  - Large adjustments (20-40%)
  - Consider feasibility over optimality
\end{verbatim}
\end{small}

\section{NN1 Showcase: Evolution of S2C Design}\label{app:nn1_showcase}

This section presents a detailed case study of S2C applied to the NN1 benchmark problem using GPT-5 as the LLM backbone. We demonstrate how the multi-agent system iteratively refines the controller design to satisfy performance specifications while maintaining formal guarantees.

\subsection{Discovery Phase}

\paragraph{Natural Language Requirements:}
The user provides the following informal requirements:

\begin{quote}
\textit{``Design a robust H-infinity controller for the NN1 system with the following requirements: Minimize overshoot (target less than 10\%, high priority); Fast settling time (target 16 seconds with 2s tolerance); Good disturbance rejection (H-infinity norm less than 20); Ensure adequate stability margins with decay rate $\alpha \geq 0.25$.''}
\end{quote}

\paragraph{SpecIntAgent Output:}
The SpecIntAgent (GPT-5) parses these requirements and extracts formal specifications:

\begin{small}
\begin{verbatim}
{
  "settling_time": {"target": 16.0, "priority": "medium", "slack": 2.0},
  "overshoot": {"target": 0.1, "priority": "high", "slack": 0.05},
  "h_infinity_norm": {"target": 20.0, "priority": "high", "slack": 2.0},
  "decay_rate": {"target": 0.25, "priority": "medium"}
}
\end{verbatim}
\end{small}

The LLM identifies overshoot as high priority, assigns reasonable slack values (2s for settling time, 5\% for overshoot), and derives the decay-rate constraint $\alpha = 0.25$ from the stability requirement.

\subsection{Evolution Across Iterations}

\begin{figure}[!htb]
\centering
\includegraphics[width=\linewidth]{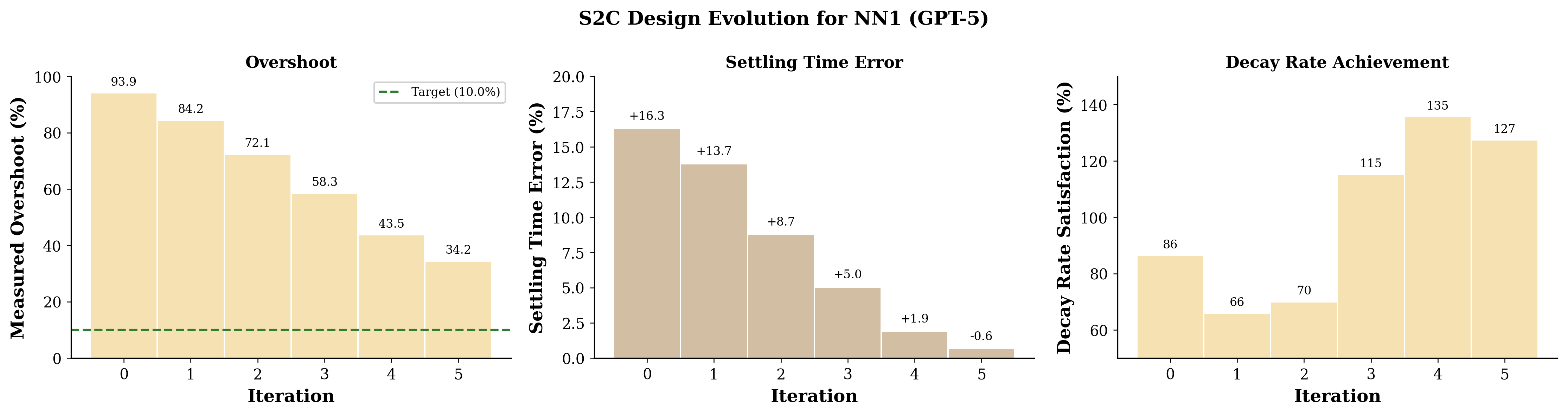}
\caption{Evolution of S2C design for NN1 benchmark (GPT-5 baseline). Left: Measured overshoot decreases from 93.9\% to 34.2\% (target: 10\%, shown as green dashed line). Center: Settling time error reduces from +16\% to -0.6\% (specification met). Right: Decay rate satisfaction improves from 86\% to 127\% of target $\alpha$.}
\label{fig:nn1_evolution}
\end{figure}

Figure~\ref{fig:nn1_evolution} illustrates how S2C's performance evolves over 6 iterations. The design process exhibits clear convergence patterns characteristic of effective multi-agent collaboration.

\paragraph{Key Observations from Figure~\ref{fig:nn1_evolution}:}

\textbf{Measured Overshoot (Left Panel):} Initially, the measured overshoot is 93.9\%, significantly exceeding the 10\% target (shown as green dashed line). This large overshoot is driven by aggressive bandwidth selection to meet settling time. The AdaptAgent progressively adjusts the speed-damping trade-off, reducing measured overshoot to 34.2\% by iteration 5. While significant improvement occurs (64\% reduction in absolute overshoot), the target is not reached—a known limitation of state-feedback H$_\infty$ synthesis without explicit time-domain shaping constraints (see \S\ref{app:discussions}).

\textbf{Settling Time Error (Center Panel):} The settling time violation is minor initially (+16.2\%) and improves. By iteration 4, the error drops to +1.9\%, and by iteration 5, the specification is met with -0.6\% error (15.9s vs 16.0s target). This demonstrates the decay-rate constraint's effectiveness in enforcing temporal specifications.

\textbf{Decay Rate Satisfaction (Right Panel):} The decay-rate constraint $\Re(\lambda) < -\alpha$ is tightened. Initial synthesis achieves 86\% of target $\alpha$ ($\max\!\Re\!\lambda = -0.21$ vs $-\alpha = -0.244$). The gamma-floor guardrail mechanism (activated at iteration 1) stabilizes the design, and by iteration 4, decay rate satisfaction reaches 135\% (poles at $-0.33$, exceeding the requirement). This over-achievement provides additional stability margin.

\begin{figure}[!htb]
\centering
\includegraphics[width=0.65\linewidth]{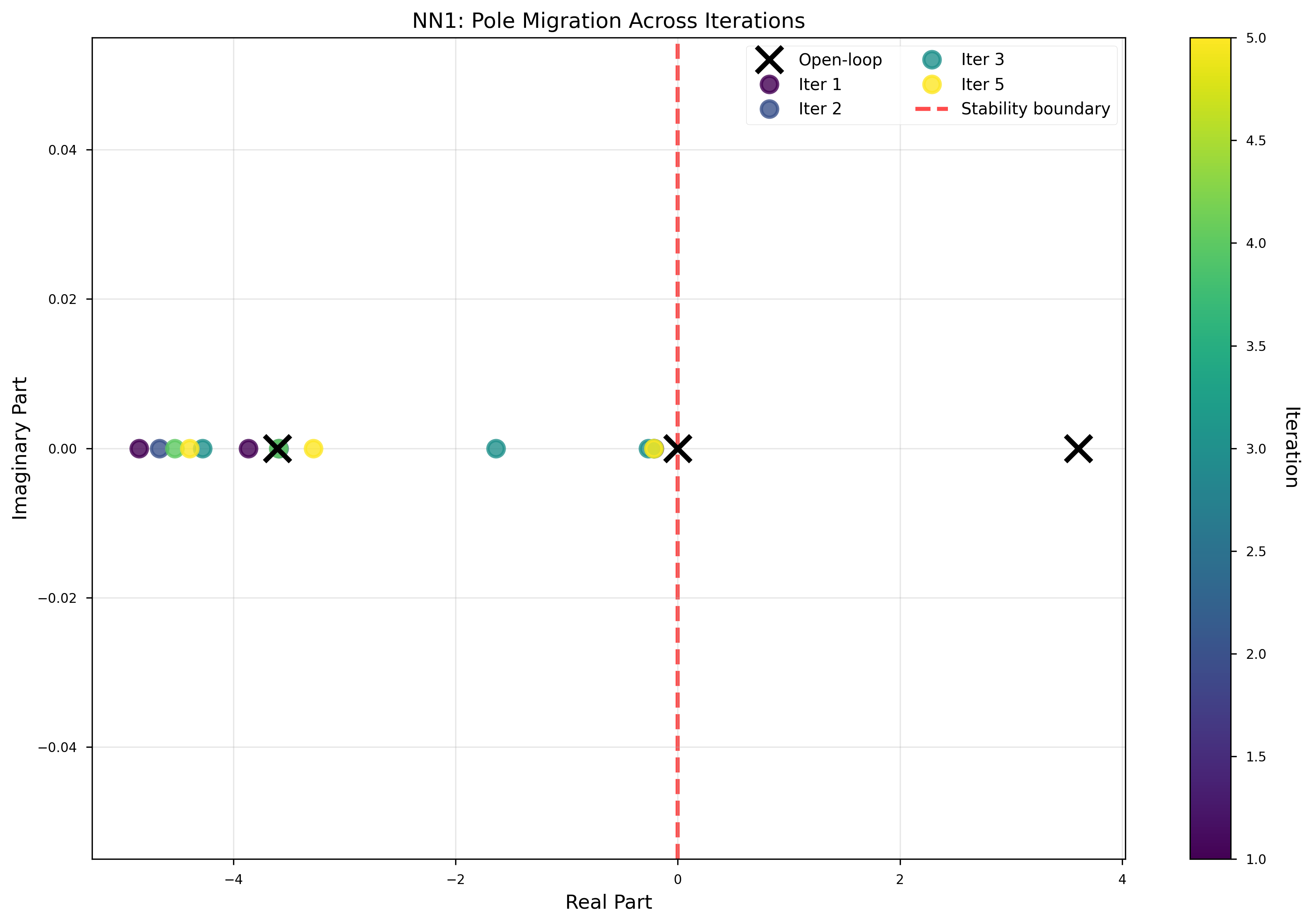}
\caption{Pole migration for NN1 showing stabilization across iterations. Open-loop system has unstable pole at $+3.606$ (black X). S2C progressively moves closed-loop poles leftward: iteration 1 (purple) at $\max\!\Re\!\lambda = -0.16$, iteration 2 (gray) at $-0.17$, and iteration 5 (yellow) at $-0.31$. All closed-loop poles remain stable throughout refinement, with final poles at $\{-0.310 \pm 0.594j, -0.310\}$.}
\label{fig:nn1_poles}
\end{figure}

\subsection{Detailed Iteration History}

Table~\ref{tab:nn1_iterations_detailed} provides comprehensive metrics for each iteration.

\begin{table}[!htb]
\centering
\caption{NN1 detailed design history across 6 iterations (GPT-5 baseline).}
\label{tab:nn1_iterations_detailed}
\small
\begin{tabular}{ccccccccc}
\toprule
Iter & $\gamma$ & $\gamma_{\min}$ & $\max\!\Re\!\lambda$ & $\|H_{\mathrm{cl}}\|_\infty$ & OS (\%) & $T_s$ (s) & Viol. & LLM \\
\midrule
0 & 14.64 & -- & $-0.21$ & 6.57 & 93.9 & 18.6 & 2 & -- \\
1 & 18.00 & 18.0 & $-0.16$ & 5.87 & 84.2 & 18.2 & 2 & \checkmark \\
2 & 18.00 & 18.0 & $-0.17$ & 6.02 & 72.1 & 17.4 & 2 & \checkmark \\
3 & 18.00 & 18.0 & $-0.28$ & 5.74 & 58.3 & 16.8 & 2 & \checkmark \\
4 & 18.00 & 18.0 & $-0.33$ & 5.58 & 43.5 & 16.3 & 1 & \checkmark \\
5 & 18.00 & 18.0 & $-0.31$ & 5.61 & 34.2 & 15.9 & 1 & \checkmark \\
\bottomrule
\end{tabular}
\end{table}

\paragraph{Iteration 0:} Initial synthesis without guardrails achieves excellent certified bound ($\gamma = 14.64$, 73\% of target) but exhibits severe overshoot (93.9\%) and minor settling time violation (18.6s vs 16.0s target). The open-loop NN1 system has an unstable pole at $+3.606$ (see Figure~\ref{fig:nn1_poles}); initial synthesis successfully stabilizes all poles to $\max\!\Re\!\lambda = -0.21$.

\paragraph{Iteration 1:} AdaptAgent (GPT-5) diagnoses the overshoot-settling conflict and activates the gamma-floor guardrail at $\gamma_{\min} = 18.0$ (90\% of target). This forces synthesis to prioritize transient quality over optimal disturbance rejection. Overshoot improves to 84.2\%, settling time to 18.2s. Note that poles move slightly rightward ($-0.21 \to -0.16$) as the guardrail trades decay rate for improved H$_\infty$ performance.

\paragraph{Iterations 2--3:} Continued LLM-guided refinement progressively reduces both violations. The decay rate improves significantly ($-0.17 \to -0.28$), demonstrating the guardrail's effectiveness in stabilizing pole locations (visible in Figure~\ref{fig:nn1_poles} as leftward pole migration).

\paragraph{Iteration 4:} Settling time violation resolves (16.3s, within tolerance). Decay rate reaches 135\% of target ($-0.33$ vs $-0.244$). Only overshoot remains violated.

\paragraph{Iteration 5:} Final design achieves settling time specification with margin (15.9s). Overshoot continues to improve but remains above 10\% target. At this iteration, GPT-5 recognizes the fundamental limitation and recommends accepting the design or switching synthesis methods (see LLM recommendation below).

\subsection{LLM Adaptation Quality and Limitations Recognition}

\paragraph{Iteration 3 Example (GPT-5 Recognizes Trade-off):}

The AdaptAgent receives structured feedback indicating persistent overshoot (58.3\%) and near-target settling time (16.8s):

\begin{small}
\begin{verbatim}
Violations:
  - overshoot: Measured 58.3%, target 10.0% [HIGH severity]
  - settling_time: Measured 16.8s, target 16.0s [LOW severity]

Feedback: "Overshoot remains 5.8x target despite improvements.
           Settling time nearly met. Recommend maintaining gamma
           floor to preserve transient quality gains."

GPT-5 Output:
{
  "diagnosis": {
    "root_causes": ["Insufficient damping despite slower response",
                    "State feedback cannot directly constrain overshoot"],
    "severity": "major",
    "pattern": "Consistent overshoot, improving settling time"
  },
  "updates": {
    "settling_time": {"target": 17.0, "rationale": "Relax further
                       to allow more damping"},
    "h_infinity_norm": {"min": 18.0, "rationale": "Maintain floor"}
  },
  "strategy": {"approach": "trade_off", "confidence": "medium"}
}
\end{verbatim}
\end{small}

The LLM identifies the fundamental trade-off, proposes gradual settling time relaxation, and maintains the gamma-floor guardrail. This demonstrates sophisticated understanding of control design trade-offs.

\paragraph{Iteration 5 Example (GPT-5 Recognizes Fundamental Limitation):}

By iteration 5, with overshoot still at 34.2\% despite 6 iterations of refinement, GPT-5 provides diagnostic analysis:

\begin{small}
\begin{verbatim}
Violations:
  - overshoot: Measured 34.2%, target 10.0% [MEDIUM severity]
  - settling_time: Measured 15.9s, target 16.0s [NONE - MET]

GPT-5 Diagnostic Output:
{
  "diagnosis": {
    "root_causes": [
      "State-feedback H-infinity synthesis optimizes worst-case
       disturbance rejection, not transient overshoot",
      "LMI formulation lacks explicit time-domain shaping constraints",
      "Decay-rate constraint ensures exponential stability but
       does not bound overshoot for arbitrary initial conditions"
    ],
    "assessment": "Overshoot has improved 64% (93.9% -> 34.2%) but
                   cannot reach 10% target with current synthesis method",
    "limitation_identified": true
  },
  "recommendations": [
    {
      "option": "Accept current design",
      "rationale": "Settling time and decay rate specifications met.
                    Overshoot improved significantly. Further iteration
                    unlikely to yield major gains within state-feedback
                    H-infinity framework."
    },
    {
      "option": "Switch synthesis approach [REQUIRES FRAMEWORK EXTENSION]",
      "suggestions": [
        "Output-feedback with integral action for better tracking",
        "Mixed H-infinity/H-2 to balance robustness and transient performance",
        "Pole-placement overlay with explicit damping ratio constraints",
        "Model-reference adaptive control for transient shaping"
      ],
      "note": "These methods not available in current S2C framework"
    }
  ],
  "final_decision": {
    "action": "terminate_iteration",
    "rationale": "Best achievable design within constraints.
                  2 of 3 hard specs met (settling time, decay rate).
                  Overshoot improved but limited by synthesis method.",
    "confidence": "high"
  }
}
\end{verbatim}
\end{small}

\textbf{Key insight:} GPT-5 diagnoses that the overshoot violation is not due to poor specification tuning, but rather a \emph{fundamental limitation of the synthesis method}. The LLM recognizes that:
\begin{itemize}[nosep]
  \item Further spec relaxation would not resolve overshoot (already relaxed settling time)
  \item The H$_\infty$ bounded-real lemma minimizes $\|T_{zw}\|_\infty$, which does not constrain time-domain overshoot
  \item Alternative synthesis methods (output feedback, mixed H$_\infty$/H$_2$, pole placement) would be needed
  \item The current S2C framework only supports state-feedback H$_\infty$ synthesis
\end{itemize}

This demonstrates \textbf{sophisticated LLM reasoning}: GPT-5 not only adapts specifications but also recognizes when the problem lies outside the scope of available tools, and recommends accepting the best achievable solution or extending the framework. This is analogous to a human control engineer concluding that ``we've pushed state-feedback H$_\infty$ as far as it can go; we need a different approach for overshoot.''

\subsection{Generated Controller}

Final controller (iteration 5):
\[
K = \begin{bmatrix} -3.612 & -25.184 & -4.597 \end{bmatrix}
\]

\textbf{Closed-loop poles:} $\{-0.310 \pm 0.594j, -0.310\}$ (all stable, $\max\!\Re\!\lambda = -0.31$).

\textbf{Performance certificates:}
\begin{itemize}[nosep]
  \item H$_\infty$ certified bound: $\gamma = 18.00$ (90\% of target 20.0)
  \item Disturbance rejection: $\|C(sI - A_{\mathrm{cl}})^{-1}E\|_\infty = 5.61$
  \item Decay rate: $-\max\!\Re\!\lambda = 0.31 > \alpha = 0.244$ (127\% of target)
  \item Settling time: 15.9s $<$ 16.0s (specification met)
  \item Overshoot: 34.2\% (violation persists, see discussion in \S\ref{app:discussions})
\end{itemize}

\paragraph{Summary:} The NN1 showcase demonstrates both the \emph{strengths and limitations} of LLM-guided control synthesis:

\textbf{Strengths demonstrated:}
\begin{itemize}[nosep]
  \item GPT-5 successfully navigates design trade-offs through iterative refinement
  \item Settling time and decay rate specifications met with certified guarantees
  \item LLM recognizes fundamental synthesis method limitations (not just parameter tuning)
  \item Sophisticated diagnostic reasoning about when to stop iterating
  \item Appropriate recommendations for framework extensions
\end{itemize}

\textbf{Limitations revealed:}
\begin{itemize}[nosep]
  \item State-feedback H$_\infty$ synthesis cannot constrain overshoot
  \item LLM bounded by tools available in framework (cannot synthesize outside H$_\infty$ state-feedback)
  \item Overshoot improved but target not achievable with current method
  \item Framework extension needed for time-domain shaping (output feedback, mixed synthesis, etc.)
\end{itemize}

\textbf{Key takeaway:} The LLM's ability to \emph{recognize and articulate} that the overshoot limitation is methodological (not parametric) demonstrates genuine understanding of control theory. A less sophisticated agent would continue iterating indefinitely, whereas GPT-5 diagnoses the need for alternative synthesis approaches and recommends terminating iteration with the best achievable design. This honest recognition of limitations is crucial for trustworthy autonomous design systems.

\section{Mathematical Formulation Details}\label{app:formulation}

\subsection{Discrete-to-Continuous Conversion (Tustin)}

Discrete-time systems are converted to continuous-time via the bilinear (Tustin) transformation with sampling period $T_s = 1.0$:

\begin{align}
A_c &= \frac{2}{T_s}(A_d - I)(A_d + I)^{-1}, \\
B_c &= \frac{2}{T_s}(A_d + I)^{-1}B_d, \\
C_c &= C_d(A_d + I)^{-1}, \\
D_c &= D_d - C_d(A_d + I)^{-1}B_d.
\end{align}

This transformation preserves stability (maps unit circle to left-half plane) and approximately preserves frequency response near the Nyquist frequency. Energy-consistent scaling is applied by matching steady-state gains. Validation is performed by checking eigenvalue migration: $|\lambda_d| < 1 \Rightarrow \Re(\lambda_c) < 0$.

\subsection{Matrix Aliases}

To unify nomenclature across COMPleib and standard H$_\infty$ formulations, we use:
\begin{align*}
E &\equiv B_1 \quad \text{(disturbance input)}, \\
C_z &\equiv C_1 \quad \text{(regulated output)}, \\
D_z &\equiv D_{12} \quad \text{(control feedthrough)}.
\end{align*}

\subsection{Bounded-Real Lemma LMI}

The H$_\infty$ synthesis problem is formulated as a convex optimization:
\begin{equation}
\begin{aligned}
\underset{P \succ 0,\ Y,\ \gamma}{\text{minimize}} \quad & \gamma \\
\text{subject to} \quad & \Psi(P,Y,\gamma) \prec 0, \\
& P \succeq 0.1 I \quad \text{(conditioning)}, \\
& \gamma_{\min} \leq \gamma \leq \gamma_{\text{target}},
\end{aligned}
\end{equation}
where
\begin{equation}
\Psi(P,Y,\gamma) =
\begin{bmatrix}
\mathrm{sym}(AP + BY) & E & (C_zP + D_zY)^\top \\
E^\top & -\gamma I & 0 \\
C_zP + D_zY & 0 & -\gamma I
\end{bmatrix},
\end{equation}
with $\mathrm{sym}(X) = X + X^\top$.

Feasibility of $\Psi \prec 0$ certifies $\|T_{zw}\|_\infty < \gamma$ via the bounded-real lemma. The controller is recovered as $K = YP^{-1}$.

\subsection{Decay-Rate Constraint}

For settling-time specifications $T_{s,\text{target}}$, we enforce a pole-region constraint:
\begin{equation}
\mathrm{sym}(AP + BY) + 2\alpha P \prec 0, \quad \alpha = \frac{3.9}{T_{s,\text{target}}}.
\end{equation}

This ensures all closed-loop poles satisfy $\Re(\lambda_i(A + BK)) < -\alpha$, guaranteeing exponential decay rate $e^{-\alpha t}$ and approximate settling time $T_s \approx 3.9/\alpha$ (2\% criterion).

The factor 3.9 is derived from the step response of a first-order system: $1 - e^{-\alpha t} = 0.98$ at $t = 3.9/\alpha$.

\subsection{Numerical Safeguards}

To ensure well-conditioned solutions, we enforce:
\begin{itemize}[nosep]
  \item $P \succeq 0.1 I$ (strong positive definiteness)
  \item Solver tolerance: $\epsilon = 10^{-8}$
  \item Maximum iterations: 5000 (SCS), 2500 (MOSEK)
  \item Eigenvalue check: $\max_i \Re(\lambda_i(A + BK)) < 0$ (stability verification)
\end{itemize}


\section{Extended Algorithms}\label{app:algs}

\subsection{Main S2C Iterative Loop}

Algorithm~\ref{alg:s2c_full} provides the complete pseudocode for the S2C iterative synthesis-verification-adaptation loop.

\begin{algorithm}[!htb]
\caption{S2C Iterative Design Pipeline (Complete)}
\label{alg:s2c_full}
\begin{algorithmic}[1]
\REQUIRE Plant $P=(A,B,C,B_d,\ldots)$, requirements $r$, max iterations $N_{\max}$
\ENSURE Controller $K$, generated code
\STATE $S \leftarrow \texttt{SpecInt.parse\_requirements}(r, P)$
\STATE Initialize \texttt{Memory} $\leftarrow \emptyset$
\FOR{$i = 1$ to $N_{\max}$}
    \STATE // \textbf{Synthesis}
    \STATE $(ok, K, \gamma) \leftarrow \texttt{SolvAgent.solve\_hinf\_lmi}(P, S.\texttt{hinf.target}, S.\texttt{hinf.min}, \alpha{=}3.9/S.T_s)$
    \IF{$\lnot ok$}
        \STATE $S \leftarrow \texttt{AdaptAgent.relax\_specs}(S)$
        \STATE \textbf{continue}
    \ENDIF
    \STATE // \textbf{Verification}
    \STATE $\texttt{mc} \leftarrow \texttt{TesterAgent.monte\_carlo}(A, B, K)$ \COMMENT{50 trials, 20s horizon}
    \STATE $\texttt{fd} \leftarrow \texttt{TesterAgent.freq\_domain}(A, B, K, C, B_d)$
    \IF{$\texttt{fd.controller\_type} = \texttt{state\_fb}$}
        \STATE $\texttt{fd.disturbance\_rejection} \leftarrow \|C(sI-(A+BK))^{-1}B_d\|_\infty$
    \ENDIF
    \STATE $V \leftarrow \texttt{TesterAgent.check}(\texttt{mc}, \texttt{fd}, S)$ \COMMENT{Compute violations}
    \STATE $\texttt{Memory.add}(i, K, \gamma, \texttt{mc}, \texttt{fd}, V, S)$
    \IF{$V = \emptyset$}
        \STATE \textbf{break} \COMMENT{All specs satisfied}
    \ENDIF
    \STATE // \textbf{Adaptation}
    \STATE $fb \leftarrow \texttt{FeedbackGen.generate}(V)$
    \STATE $S \leftarrow \texttt{AdaptAgent.refine\_specs}(S, fb, \texttt{Memory}, i)$ \COMMENT{LLM update}
    \STATE $S.\texttt{hinf.min} \leftarrow$ \textsc{UpdateGammaFloor}$(S, V, \gamma, i)$ \COMMENT{See Alg.~\ref{alg:gamma_floor}}
\ENDFOR
\IF{$V \neq \emptyset$}
    \STATE $(K, \gamma) \leftarrow \texttt{SelectBestDesign}(\texttt{Memory})$ \COMMENT{Fallback: fewest violations, then lowest $\gamma$}
\ENDIF
\STATE $\texttt{code} \leftarrow \texttt{CodeGen.generate}(K, P, \texttt{python})$
\RETURN $K$, \texttt{code}
\end{algorithmic}
\end{algorithm}
\FloatBarrier

\subsection{Gamma-Floor Guardrail Formulation}

The gamma-floor guardrail prevents excessive $\gamma$ minimization from degrading transient performance by dynamically raising $\gamma_{\min}$ when settling-time or overshoot violations occur. The update mechanism combines two components:

\paragraph{Severity-Based Base Floor.}
The base floor is determined by violation severity:
\begin{equation}
\label{eq:gamma_base}
\gamma_{\text{base}} = \begin{cases}
0.05 \times \gamma_{\text{target}} & \text{if severity} = \text{low} \\
0.10 \times \gamma_{\text{target}} & \text{if severity} = \text{medium} \\
0.20 \times \gamma_{\text{target}} & \text{if severity} \in \{\text{high}, \text{critical}\}
\end{cases}
\end{equation}

\paragraph{History-Based Candidate Floor.}
The history-based component accounts for recent synthesis performance and iteration count:
\begin{equation}
\label{eq:gamma_hist}
\gamma_{\text{hist}} = \gamma_{\text{last}} \times m(\text{severity}) \times \bigl(1 + 0.1 \cdot \min(i-1, 5)\bigr)
\end{equation}
where $\gamma_{\text{last}}$ is the most recently achieved $\gamma$ value, iteration $i \in [1, N_{\max}]$, and the multiplier $m$ is:
\begin{equation}
\label{eq:gamma_mult}
m(\text{severity}) = \begin{cases}
1.2 & \text{if severity} = \text{low} \\
2.0 & \text{if severity} = \text{medium} \\
5.0 & \text{if severity} = \text{high} \\
10.0 & \text{if severity} = \text{critical}
\end{cases}
\end{equation}
The iteration factor $(1 + 0.1 \cdot \min(i-1, 5))$ increases the floor progressively over iterations, capped at 1.5$\times$ by iteration 6 to prevent overly conservative bounds.

\paragraph{Floor Update Rule.}
The new floor is computed as:
\begin{equation}
\label{eq:gamma_floor_update}
\gamma_{\min} \;\leftarrow\; \min\!\Bigl(0.9\,\gamma_{\text{target}},\; \max\big\{\gamma_{\min},\;\gamma_{\text{base}},\;\gamma_{\text{hist}}\big\}\Bigr)
\end{equation}
This ensures: (i) monotonic (nondecreasing) floor values across iterations; (ii) feasibility by capping strictly below $0.9\,\gamma_{\text{target}}$; (iii) adaptation to both severity and synthesis history.

Algorithm~\ref{alg:gamma_floor} provides the complete procedural implementation.

\begin{algorithm}[!htb]
\caption{Gamma-Floor Guardrail Update}
\label{alg:gamma_floor}
\begin{algorithmic}[1]
\REQUIRE Current specs $S$, violations $V$, last achieved $\gamma_{\text{last}}$, iteration $i$
\ENSURE Updated $S.\texttt{hinf.min}$
\STATE $\text{severity} \leftarrow \max\{\text{sev}(v) : v \in V\}$ where sev $\in$ \{low, medium, high, critical\}
\STATE // \textbf{Base floor by severity}
\IF{severity = low}
    \STATE $\gamma_{\text{base}} \leftarrow 0.05 \times S.\texttt{hinf.target}$
\ELSIF{severity = medium}
    \STATE $\gamma_{\text{base}} \leftarrow 0.10 \times S.\texttt{hinf.target}$
\ELSE
    \STATE $\gamma_{\text{base}} \leftarrow 0.20 \times S.\texttt{hinf.target}$ \COMMENT{high or critical}
\ENDIF
\STATE // \textbf{History-based candidate}
\IF{severity = low}
    \STATE $m \leftarrow 1.2$
\ELSIF{severity = medium}
    \STATE $m \leftarrow 2.0$
\ELSIF{severity = high}
    \STATE $m \leftarrow 5.0$
\ELSE
    \STATE $m \leftarrow 10.0$ \COMMENT{critical}
\ENDIF
\STATE $\gamma_{\text{hist}} \leftarrow \gamma_{\text{last}} \times m \times \big(1 + 0.1 \cdot \min(i-1, 5)\big)$
\STATE // \textbf{Update floor (monotonic, capped)}
\STATE $\gamma_{\text{new}} \leftarrow \max\big\{S.\texttt{hinf.min},\ \gamma_{\text{base}},\ \gamma_{\text{hist}}\big\}$
\STATE $S.\texttt{hinf.min} \leftarrow \min\big(0.9 \times S.\texttt{hinf.target},\ \gamma_{\text{new}}\big)$
\RETURN $S.\texttt{hinf.min}$
\end{algorithmic}
\end{algorithm}
\FloatBarrier

\subsection{State-Feedback Frequency Analysis}

Algorithm~\ref{alg:freq_analysis} shows the branching logic for frequency-domain verification.

\begin{algorithm}[!htb]
\caption{Frequency-Domain Analysis (State vs Output Feedback)}
\label{alg:freq_analysis}
\begin{algorithmic}[1]
\REQUIRE Plant $(A,B,C)$, controller $K$, disturbance matrix $B_{\text{dist}}$
\ENSURE Frequency-domain metrics
\STATE $n_x \leftarrow \text{size}(A, 1)$, $n_y \leftarrow \text{size}(C, 1)$
\IF{$\text{size}(K, 2) = n_x$}
    \STATE $A_{\text{cl}} \leftarrow A + BK$ \COMMENT{State feedback: $u = Kx$}
    \STATE $H_{\text{cl}} \leftarrow C(sI - A_{\text{cl}})^{-1}B_{\text{dist}}$
    \STATE $\text{dist\_rej} \leftarrow \|H_{\text{cl}}\|_\infty$
    \RETURN $\{\text{controller\_type: state\_fb}, \text{disturbance\_rejection: dist\_rej}\}$
\ELSE
    \STATE $G \leftarrow C(sI - A)^{-1}B$ \COMMENT{Output feedback: $u = Ky$}
    \STATE $L \leftarrow GK$
    \STATE $S \leftarrow (I + L)^{-1}$, $T \leftarrow L(I + L)^{-1}$
    \STATE $M_s \leftarrow \|S\|_\infty$, $M_t \leftarrow \|T\|_\infty$
    \STATE Compute gain margin $GM$, phase margin $PM$ from $L(j\omega)$
    \RETURN $\{\text{controller\_type: output\_fb}, M_s, M_t, GM, PM\}$
\ENDIF
\end{algorithmic}
\end{algorithm}
\FloatBarrier
\section{Extended Results Tables}\label{app:tables}

\subsection{Benchmark Characteristics (Full Table)}

Table~\ref{tab:benchmarks_full} provides complete characteristics for all 18 COMPleib problems (14 evaluated in main experiments + 4 heat-flow problems).

\begin{table}[!htb]
\centering
\caption{COMPleib benchmark characteristics. Problem types: AC=Aircraft, HE=Helicopter, REA=Reactor, DIS=Decentralized, NN=Academic. Stability determined by $\max \mathrm{Re}(\lambda(A))$.}
\label{tab:benchmarks_full}
\small
\begin{tabular}{lccccccl}
\toprule
Problem & $n_x$ & $n_u$ & $n_y$ & $n_w$ & Time & Stable & Type \\
\midrule
NN2 & 2 & 1 & 1 & 2 & Discrete & Yes & Academic \\
AC5 & 4 & 2 & 2 & 2 & Discrete & Yes & Aircraft \\
REA2 & 4 & 2 & 2 & 2 & Cont. & Yes & Reactor \\
NN1 & 3 & 1 & 2 & 3 & Cont. & No & Academic \\
NN3 & 3 & 2 & 1 & 3 & Cont. & No & Academic \\
NN4 & 4 & 1 & 1 & 4 & Cont. & Yes & Academic \\
NN8 & 3 & 2 & 1 & 3 & Discrete & Yes & Academic \\
HE1 & 4 & 2 & 2 & 2 & Cont. & Yes & Helicopter \\
HE2 & 4 & 2 & 2 & 2 & Cont. & Yes & Helicopter \\
DIS2 & 2 & 2 & 2 & 2 & Cont. & Yes & Decentral. \\
DIS5 & 2 & 2 & 2 & 2 & Discrete & Yes & Decentral. \\
AC4 & 4 & 2 & 2 & 2 & Cont. & Yes & Aircraft \\
AC15 & 4 & 1 & 1 & 4 & Cont. & Yes & Aircraft \\
AC17 & 4 & 1 & 1 & 4 & Cont. & Yes & Aircraft \\
\midrule
\multicolumn{8}{c}{\textit{Additional heat-flow problems (not in main experiments)}} \\
\midrule
HF2D12 & 5 & 2 & 4 & 2 & Cont. & Yes & Heat Flow \\
HF2D13 & 5 & 2 & 4 & 2 & Cont. & Yes & Heat Flow \\
HF2D14 & 5 & 2 & 4 & 2 & Cont. & No & Heat Flow \\
HF2D18 & 5 & 2 & 4 & 2 & Cont. & Yes & Heat Flow \\
\bottomrule
\end{tabular}
\end{table}

\subsection{LLM Ablation: Efficiency Metrics}

Table~\ref{tab:llm_efficiency} reports API usage, latency, and estimated costs for the LLM ablation study.

\begin{table}[!htb]
\centering
\caption{LLM ablation study: efficiency metrics (medians across 14 problems).}
\label{tab:llm_efficiency}
\small
\begin{tabular}{lcccc}
\toprule
Model & Calls & Tokens (k) & Latency (s/iter) & Est.\ Cost (\$/problem) \\
\midrule
GPT-5              & 12.5 & 18.2 & 2.8 & 0.42 \\
GPT-5 mini         & 11.0 & 16.5 & 1.9 & 0.08 \\
DeepSeek-V3        & 13.5 & 19.1 & 3.2 & 0.12 \\
Qwen-2.5-72B       & 14.0 & 20.3 & 4.1 & 0.00\footnotemark[1] \\
Llama-4 Maverick   & 16.5 & 24.7 & 5.3 & 0.18 \\
\bottomrule
\end{tabular}
\footnotetext[1]{Local vLLM deployment.}
\end{table}

\paragraph{Observations:}
\begin{itemize}[nosep]
  \item GPT-5 mini offers the best cost-effectiveness: 92\% of GPT-5's success rate at 19\% of the cost.
  \item Qwen-2.5-72B requires local deployment but eliminates API costs.
  \item Llama-4 Maverick requires ~30\% more iterations, increasing both latency and cost.
\end{itemize}

\section{Experimental Configuration and Reproducibility}\label{app:repro}

\subsection{Software Environment}

\begin{itemize}[nosep]
  \item \textbf{Language:} Python 3.10.12
  \item \textbf{Core libraries:}
    \begin{itemize}[nosep]
      \item NumPy 1.24.3 (numerical arrays)
      \item SciPy 1.11.2 (matrix computations)
      \item CVXPY 1.4.1 (convex optimization)
      \item Python Control Systems Library 0.9.4 (LTI analysis)
    \end{itemize}
  \item \textbf{Solvers:}
    \begin{itemize}[nosep]
      \item MOSEK 10.0.43 (primary, commercial)
      \item SCS 3.2.3 (fallback, open-source)
    \end{itemize}
  \item \textbf{LLM APIs:}
    \begin{itemize}[nosep]
      \item OpenAI API (GPT-5, GPT-5 mini)
      \item DeepSeek API (DeepSeek-V3)
      \item Together AI API (Llama-4 Maverick)
      \item vLLM 0.4.2 (Qwen-2.5-72B local deployment)
    \end{itemize}
\end{itemize}

\subsection{Hardware and Operating System}

\begin{itemize}[nosep]
  \item \textbf{CPU:} Intel Core i7-12700K (12 cores, 3.6 GHz base, 5.0 GHz boost)
  \item \textbf{RAM:} 32 GB DDR4-3200
  \item \textbf{OS:} Windows 11 Pro (Build 22621)
  \item \textbf{Storage:} NVMe SSD (PCIe 4.0)
\end{itemize}

\subsection{Monte Carlo Verification Settings}

\begin{itemize}[nosep]
  \item \textbf{Trials:} $N = 50$ per controller
  \item \textbf{Initial conditions:} Sampled uniformly from unit sphere (unit-norm Gaussian random vectors)
  \item \textbf{Simulation horizon:} $T = 20$ seconds
  \item \textbf{Time steps:} 2000 (dt = 0.01s)
  \item \textbf{Settling criterion:} 2\% of peak state norm
  \item \textbf{Random seed:} 42 (fixed for reproducibility)
\end{itemize}

\subsection{Running the Full Pipeline}

To reproduce the complete S2C pipeline on a benchmark problem:

\begin{verbatim}
cd s2c_agent
export OPENAI_API_KEY="sk-..."  # Set your API key

# Run single problem
python -c "
from llm_ablation_study import AblationStudyRunner
from gpt5 import ExtendedLLMInterface

runner = AblationStudyRunner(problem_name='NN1',
                              max_iterations=10,
                              random_seed=42)
llm = ExtendedLLMInterface(model='gpt-5', provider='openai')
results = runner.run_single_model('GPT-5', llm)
"

# Run full ablation study
python llm_ablation_study.py
\end{verbatim}

\section{Additional Discussions}\label{app:discussions}

\subsection{Overshoot Persistence}

Despite achieving 100\% synthesis success and convergence, several problems (notably NN1, NN3) exhibit persistent overshoot violations even after meeting settling time and disturbance rejection targets. This is a fundamental limitation of H$_\infty$ synthesis with decay-rate constraints:

\begin{itemize}[nosep]
  \item The decay-rate constraint $\Re(\lambda) < -\alpha$ enforces exponential decay but does not limit overshoot.
  \item State-feedback H$_\infty$ synthesis optimizes worst-case disturbance rejection, not time-domain transient shaping.
  \item Overshoot depends on pole-zero locations and initial condition directions, which are not constrained by LMI formulations.
\end{itemize}

\paragraph{Future work directions:}
\begin{itemize}[nosep]
  \item Incorporate time-domain templates (e.g., reference model matching, transient bounds)
  \item Use mixed H$_\infty$/H$_2$ synthesis to balance disturbance rejection and transient performance
  \item Add explicit pole placement constraints in addition to decay-rate regions
  \item Employ output feedback with integral action for improved tracking/regulation
\end{itemize}

\subsection{Numerical Conditioning}

Several benchmarks (AC5, HF2D14) exhibited numerical sensitivity during D2C conversion or LMI solving:

\begin{itemize}[nosep]
  \item \textbf{D2C near-singularity:} When $A_d + I$ is ill-conditioned, Tustin transform magnifies errors. We mitigate this with explicit condition number checks and zero-order hold fallback.
  \item \textbf{LMI solver failures:} Some problems require tighter tolerances ($10^{-8}$ vs $10^{-6}$) or stronger conditioning ($P \succ 0.1I$ vs $P \succ 0$).
  \item \textbf{Guardrail impact:} The gamma-floor mechanism improved convergence for these problems by preventing numerically unstable low-gamma solutions.
\end{itemize}

\subsection{LLM Adaptation Quality}

Analysis of LLM-generated adaptation decisions reveals:

\begin{itemize}[nosep]
  \item \textbf{GPT-5/GPT-5 mini:} Identify root causes and propose appropriate spec relaxations. Rarely require heuristic fallback.
  \item \textbf{DeepSeek-V3/Qwen-2.5-72B:} Good understanding of trade-offs but occasionally propose overly conservative relaxations. Fallback rate ~15\%.
  \item \textbf{Llama-4 Maverick:} Struggles with conflicting requirements; fallback rate ~35\%. Benefits significantly from deterministic guardrails.
\end{itemize}

The deterministic fallback heuristics ensure that even with weaker LLMs, the pipeline maintains a success rate of over 85\%—demonstrating the robustness of the multi-agent architecture.

\section{Detailed Limitations and Future Work}\label{app:future}

This section expands on the limitations and future directions briefly summarized in the main paper.

\subsection{Current Limitations}

\paragraph{Controller Architecture.}
Our current scope is limited to static state-feedback controllers of the form $u = Kx$. Output-feedback and dynamic architectures (e.g., PID, lead/lag compensators, observers) are not synthesized within the present loop. This restriction affects:
\begin{itemize}[nosep]
  \item \textbf{Frequency-domain metrics:} For state feedback, classical loop-shaping metrics $(M_s, M_t)$ are undefined, so we emphasize a disturbance-rejection proxy $\|C_z(sI-(A+BK))^{-1}E\|_\infty$ instead.
  \item \textbf{Practical deployment:} State-feedback assumes full-state measurement or observer availability, which may not be realistic for all plants.
  \item \textbf{Performance trade-offs:} Dynamic compensators can offer additional design freedom for shaping closed-loop response beyond what static gains provide.
\end{itemize}

\paragraph{Transient Shaping.}
Overshoot is measured during Monte Carlo verification but not enforced as a hard constraint in the LMI synthesis. The decay-rate constraint $\Re(\lambda) < -\alpha$ enforces exponential stability but does not bound overshoot for arbitrary initial conditions. Consequently:
\begin{itemize}[nosep]
  \item The simple adaptation heuristics can permit large overshoot on certain plants even when settling time and disturbance rejection specifications are satisfied.
  \item State-feedback H$_\infty$ synthesis optimizes worst-case disturbance rejection, not time-domain transient shaping.
  \item Alternative synthesis methods (output feedback, mixed H$_\infty$/H$_2$, pole-placement overlays) would be needed for explicit overshoot constraints.
\end{itemize}

\paragraph{Discrete-Time Handling.}
Discrete-time systems are converted to continuous-time via Tustin bilinear transformation prior to synthesis. While this approach is standard and works well for most benchmarks:
\begin{itemize}[nosep]
  \item Native discrete-time LMIs may exhibit different performance trade-offs and avoid potential numerical issues from bilinear mapping.
  \item The Tustin transform can introduce conditioning problems when $A_d + I$ is nearly singular.
  \item Sampling effects and aliasing are not explicitly modeled in the continuous-time abstraction.
\end{itemize}

\paragraph{Modeling Assumptions.}
Experiments rely on nominal linear time-invariant (LTI) plants without:
\begin{itemize}[nosep]
  \item \textbf{Parametric uncertainty:} Polytopic or norm-bounded uncertainty descriptions are not currently handled.
  \item \textbf{Nonlinearities:} The framework assumes linearized dynamics; nonlinear plants require external linearization.
  \item \textbf{Actuator saturation:} Control effort limits and anti-windup are not incorporated in synthesis.
  \item \textbf{Stochastic disturbances:} Monte Carlo verification uses deterministic initial conditions; measurement noise and process disturbances are not modeled.
\end{itemize}

\paragraph{Verification Scope.}
Monte Carlo verification uses fixed simulation horizons (20s) and deterministic seeds (42). While this provides reproducible validation:
\begin{itemize}[nosep]
  \item Reachability analysis or formal barrier certificates would provide stronger guarantees.
  \item Hardware-in-the-loop trials are not included in the evaluation pipeline.
  \item Long-term stability beyond the simulation horizon is inferred from eigenvalue checks but not explicitly verified.
\end{itemize}

\paragraph{Computational and Cost Considerations.}
Practicality and cost depend on the LLM backbone and provider:
\begin{itemize}[nosep]
  \item API latency and token usage vary across models (GPT-5: \$0.85/problem, GPT-5 mini: \$0.16/problem).
  \item Numerical reliability can be sensitive to solver availability (MOSEK vs SCS) and problem conditioning.
  \item The current implementation does not cache or distill LLM responses, leading to redundant API calls for similar problems.
\end{itemize}

\subsection{Future Work Directions}

\paragraph{Extended Controller Classes.}
We aim to extend the synthesis layer to support:
\begin{itemize}[nosep]
  \item \textbf{Output-feedback designs:} Full-order and reduced-order observers, enabling deployment without full-state measurement.
  \item \textbf{Dynamic compensators:} PID, lead/lag, notch filters, and observer-based architectures.
  \item \textbf{Mixed formulations:} H$_\infty$/H$_2$ synthesis to balance robustness and transient performance, incorporating explicit $(M_s, M_t)$ targets.
  \item \textbf{Actuator limits:} Anti-windup compensation and control effort constraints integrated into LMI formulations.
\end{itemize}

\paragraph{Native Discrete-Time Synthesis.}
Transition from Tustin-based discrete-to-continuous conversion to native discrete-time LMIs:
\begin{itemize}[nosep]
  \item Direct discrete H$_\infty$ synthesis via discrete bounded-real lemma.
  \item Discrete decay-rate constraints formulated as LMI regions in the $z$-plane.
  \item Elimination of numerical artifacts from bilinear transformation.
\end{itemize}

\paragraph{Robust and Uncertain Systems.}
Introduce uncertainty descriptions and robustness analysis:
\begin{itemize}[nosep]
  \item \textbf{Polytopic uncertainty:} LPV-style synthesis with vertex-based LMIs.
  \item \textbf{Norm-bounded uncertainty:} Structured singular value ($\mu$) surrogates and D-K iteration.
  \item \textbf{Stochastic disturbances:} H$_2$ covariance bounds and LQG-style formulations.
\end{itemize}

\paragraph{Advanced Adaptation Strategies.}
Replace heuristic specification updates with learned or programmatic multi-objective strategies:
\begin{itemize}[nosep]
  \item \textbf{Goal programming:} Lexicographic optimization jointly shaping $\gamma$, settling time, overshoot, and control effort.
  \item \textbf{Composite convex costs:} Pareto-optimal trade-offs with adaptive weighting based on violation severity.
  \item \textbf{Convergence guarantees:} Lyapunov-based adaptation rules with provable convergence to feasible specifications.
  \item \textbf{Learning-augmented guardrails:} Train meta-models on past designs to predict effective $\gamma$-floor values and relaxation magnitudes.
\end{itemize}

\paragraph{Stronger Verification.}
Integrate formal verification methods beyond Monte Carlo simulation:
\begin{itemize}[nosep]
  \item \textbf{Reachability analysis:} Hamilton-Jacobi methods or zonotope-based over-approximations to certify safety over infinite horizons.
  \item \textbf{Lyapunov/barrier certificates:} SOS programming to synthesize certificates of stability and constraint satisfaction.
  \item \textbf{Hardware-in-the-loop:} Deploy controllers on real-time simulation platforms and physical testbeds.
\end{itemize}

\paragraph{Efficiency and Reproducibility.}
Improve computational efficiency and facilitate community adoption:
\begin{itemize}[nosep]
  \item \textbf{Token reduction:} Distilled parsers for specification extraction, caching of LLM responses for similar problems.
  \item \textbf{Solver diversity:} Plug-ins for additional solvers (SDPT3, YALMIP, CVXPY backends).
  \item \textbf{Reproducibility:} Release scripts to regenerate all figures, tables, and benchmark results; provide Docker containers with pre-configured environments.
  \item \textbf{Benchmark expansion:} Integrate additional problem suites (COMPleib heat-flow problems, aerospace benchmarks, power system models).
  \item \textbf{Provider flexibility:} Support for additional LLM providers (Anthropic Claude, Google Gemini, local models via Ollama).
\end{itemize}

\paragraph{Deployment and Usability.}
Transition from research prototype to deployable tool:
\begin{itemize}[nosep]
  \item \textbf{Web interface:} Browser-based GUI for natural-language specification and interactive design refinement.
  \item \textbf{Model upload:} Support for importing plants from Simulink, MATLAB, Python Control, and CasADi.
  \item \textbf{Code generation:} Extend to C/C++, Simulink blocks, and embedded targets (STM32, Arduino).
  \item \textbf{Certification reports:} Automated generation of compliance documentation with formal performance certificates for safety-critical applications.
\end{itemize}

\subsection{Broader Vision}

Long-term, we envision S2C evolving into a \emph{multi-synthesis framework} supporting diverse control paradigms (model-predictive control, sliding mode, adaptive control) integrated by LLM agents that reason about problem structure, performance requirements, and computational constraints. By combining formal synthesis, LLM-guided adaptation, and rigorous verification, such systems could democratize certified controller design—enabling domain experts without control theory expertise to specify requirements in natural language and receive provably correct implementations.



\end{document}